\documentclass[12pt]{iopart}

\usepackage{graphicx}
\usepackage{subfig}
\usepackage{amssymb}
\usepackage{amsthm}
\usepackage{mathptmx}
\usepackage{nicefrac}
\usepackage{float}
\usepackage{xcolor}
\usepackage{multirow}
\usepackage{multicol}
\usepackage{textgreek}
\usepackage[switch]{lineno} 

\begin{document}
\title[]{Extended X-ray energy characterization of SIDDHARTA-2 large-area Silicon Drift Detectors up to 50 keV}

\author{ 
F Clozza$^{1,2,*}$, F Sgaramella$^{1}$, L Abbene$^{3,1}$, F Artibani$^{1,4}$, M Bazzi$^{1}$, G Borghi$^{5,6}$, D Bosnar$^{7,1}$, M Bragadireanu$^{8}$, A Buttacavoli$^{3,1}$, M Carminati$^{5,6}$, A Clozza$^1$, R Del Grande$^{9,1}$, L De Paolis$^1$, K Dulski$^{1,10,11}$, F Ficorella$^{12}$, C Fiorini$^{5,6}$, I Fri\v{s}\v{c}i\'c$^{7}$, C Guaraldo$^{1,^\dagger}$, M Iliescu$^1$, M Iwasaki$^{13}$, A Khreptak$^{10,11,1}$, S Manti$^1$, J Marton$^{14}$, P Moskal$^{10,11}$, F Napolitano$^{15,16,1}$, S Nied\'{z}wiecki$^{10,11,1}$, H Ohnishi$^{17}$, A Picciotto$^{12}$, K Piscicchia$^{18,1}$, F Principato$^{3,1}$, A Scordo$^{1}$, M Silarski$^{10}$, D Sirghi$^{18,1,8}$, F Sirghi$^{1,8}$, M Skurzok$^{10,11,1}$, A Spallone$^1$, K Toho$^{17,1}$, O Vazquez Doce$^1$, N Zorzi$^{12}$, J Zmeskal$^{19,1,^\dagger}$ and C Curceanu$^1$}

\address{$^1$ Laboratori Nazionali di Frascati INFN, Frascati, Italy}
\address{$^2$ Università degli studi di Roma Tor Vergata, Dipartimento di Fisica, Roma, Italy}
\address{$^3$ Department of Physics and Chemistry (DiFC)—Emilio Segrè, University of Palermo, Palermo, Italy}
\address{$^4$ Università degli studi di Roma Tre, Dipartimento di Fisica, Roma, Italy}
\address{$^5$ Politecnico di Milano, Dipartimento di Elettronica, Informazione e Bioingegneria, Milano, Italy}
\address{$^6$ INFN Sezione di Milano, Milano, Italy}
\address{$^7$ Department of Physics, Faculty of Science, University of Zagreb, Zagreb, Croatia}
\address{$^8$ Horia Hulubei National Institute of Physics and Nuclear Engineering (IFIN-HH) Măgurele, Romania}
\address{$^9$ Faculty of Nuclear Sciences and Physical Engineering, Czech Technical University in Prague, Břehová 7, 115 19, Prague, Czech Republic}
\address{$^{10}$ Faculty of Physics, Astronomy, and Applied Computer Science, Jagiellonian University, Krakow, Poland}
\address{$^{11}$ Center for Theranostics, Jagiellonian University, Krakow, Poland}
\address{$^{12}$  Sensors $\&$ Devices Center - Fondazione Bruno Kessler, Via Sommarive 18, Trento, 38123, Italy}
\address{$^{13}$ RIKEN, Tokyo, Japan}
\address{$^{14}$ Atominstitut, Technical University Vienna, Stadionallee 2, 1020, Vienna, Austria}
\address{$^{15}$ Via A. Pascoli 06123, Perugia (PG), Italy, Dipartimento di Fisica e Geologia, Università degli studi di Perugia}
\address{$^{16}$ INFN Sezione di Perugia, Via A. Pascoli – 06123 Perugia – Italia}
\address{$^{17}$ Research Center for Accelerator and Radioisotope Science (RARiS), Tohoku University, Sendai, Japan}
\address{$^{18}$ Centro Ricerche Enrico Fermi – Museo Storico della Fisica e Centro Studi e Ricerche “Enrico Fermi”, Roma, Italy}
\address{$^{19}$ Stefan-Meyer-Institut f\"ur Subatomare Physik, Vienna, Austria}
\address{$^\dagger$ deceased}

\ead{$^*$ francesco.clozza@lnf.infn.it (Corresponding Author)}

\vspace{10pt}

\begin{abstract}
The SIDDHARTA-2 experiment at the DA$\Phi$NE collider of INFN-LNF performs high precision light kaonic atoms X-ray spectroscopy to investigate the kaon-nucleon(s) strong interaction in the low-energy (O(10 keV)) regime. A large area Silicon Drift Detectors (SDDs) system has been developed to carry out these measurements. The collaboration aims to extend the measurements campaign to higher mass kaonic atoms, which exhibit transition lines at increased X-ray energies. In this context, the spectroscopic response of the SIDDHARTA-2 SDD system was investigated in terms of linearity and energy resolution up to 50 keV. An accuracy of the energy calibration procedure $\Delta E/E < 10^{-3}$ was achieved.
\end{abstract}

\vspace{2pc}

\section{Introduction}
Thanks to their excellent performance in terms of energy and time resolutions and to their high rate capability, the Silicon Drift Detector (SDD) have become a key instrument in high precision X-ray spectroscopy experiments. In particular, SDDs are ideal detectors to perform kaonic atoms X-ray spectroscopy \cite{PhysRev.58.90.2,Bazzi:2011zza}, used to extract information on the low-energy strong interaction in antikaon-nucleon(s) systems. This information can be obtained by measuring, through X-ray spectroscopy, the shift ($\varepsilon$), with respect to the predictions of Quantum Electrodynamics alone, and the broadening (the width, $\Gamma$) of the atomic energy levels caused by the antikaon-nucleon(s) strong interaction \cite{Curceanu:2026zjg}. The SIDDHARTA-2 experiment \cite{Curceanu:2019uph,Sirghi:2023wok,Artibani:2024kop} at the DA$\Phi$NE collider of Laboratori Nazionali di Frascati of INFN (INFN-LNF) \cite{Milardi:2018sih,Milardi:2021khj,Milardi:2024efr} employs an SDDs based system \cite{Miliucci:2021wbj,Miliucci:2022lvn} for precision X-ray measurements of kaonic atoms. In 2021, motivated by the primary goal of the SIDDHARTA-2 collaboration to perform the first measurement of kaonic deuterium \cite{Curceanu:2019uph}, the energy response of the SDD system was characterized in the 4–12 keV energy range, corresponding to the region of interest for this measurement. These studies demonstrated a highly linear energy response, at the level of a few eV (\textit{i.e.} $\Delta E/E < 10^{-3}$), and an energy resolution (FWHM) of approximately 160 eV at 6.4 keV \cite{Miliucci:2021wbj}.\\
The SIDDHARTA-2 collaboration is currently exploring the feasibility of new measurements of intermediate-mass and heavy kaonic atoms, motivated by their potential impact on precision studies ranging from QED to low-energy QCD in the strangeness sector \cite{Curceanu:2026zjg,Curceanu:2023yuy}. Kaonic lithium, beryllium, and boron are particularly suitable for these investigations, as their strong interaction induced shifts and widths provide access to the study of kaon–multi-nucleon interactions \cite{Curceanu:2026zjg}, while simultaneously offering stringent tests of bound-state QED in few-body systems \cite{Paul:2020cnx,rry5-tdqb}. Their X-ray transitions lie in the 15–40 keV energy range, with expected strong-interaction widths spanning from 40 to 800 eV. In this context, a detailed characterization of the spectroscopic response of the SIDDHARTA-2 Silicon Drift Detector (SDD) system over an extended energy range is mandatory. To this end, a dedicated study of the SDD energy response from 10 to 50 keV has been performed for the first time.

\section{The SIDDHARTA-2's Silicon Drift Detectors}
\begin{figure}[H]
    \centering
    \includegraphics[width=0.50\textwidth]{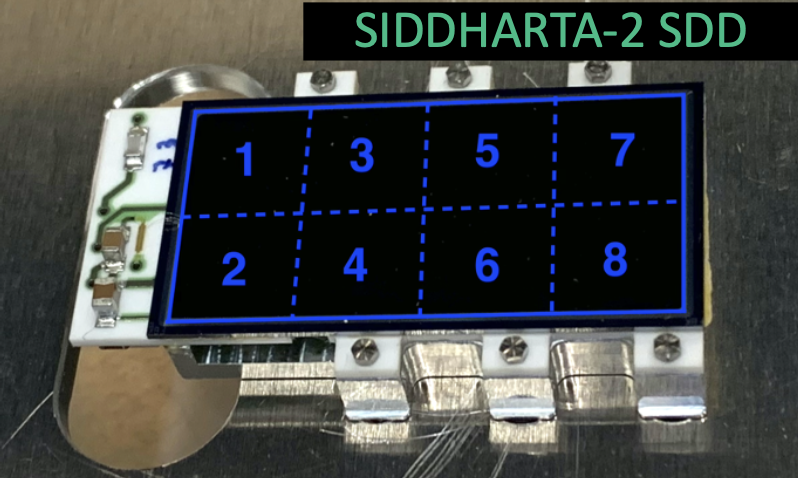}
    \includegraphics[width=0.70\textwidth]{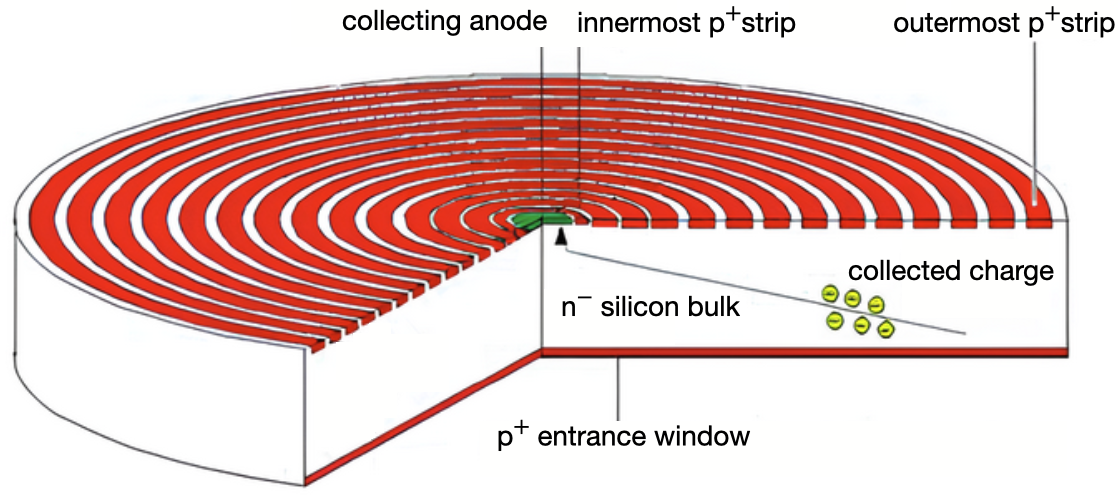}
    \caption{Above: Picture of an SDDs array used by the SIDDHARTA-2 experiment. Below: schematic representation of the SDD and of the charge collection process.}
    \label{fig:sdds}
\end{figure}

The SIDDHARTA-2 experiment employs an X-ray spectroscopy system based on large-area monolithic arrays of Silicon Drift Detectors (SDDs). The system consists of 48 arrays for a total of 384 SDD units, covering an active area of 245 cm$^2$, and is arranged around the SIDDHARTA-2 cryogenic target cell to detect X-rays from the de-excitation of kaonic atoms produced in the target \cite{Sirghi:2023wok}. The SDD, shown in Figure \ref{fig:sdds}, consists of a $n^-$ silicon substrate with ring-shaped $p^+$ implants on one side and a shallow $p^+$ entrance window on the other, providing homogeneous sensitivity over the detector area \cite{lechner1996silicon,lechner2001silicon,lechner2004novel}. Each monolithic SDD array (Figure \ref{fig:sdds}), developed by Fondazione Bruno Kessler (FBK, Italy) in collaboration with the Politecnico di Milano (PoliMi, Italy), INFN-LNF and Stefan Meyer Institute (SMI, Austria), consists of eight squared SDD cells, arranged into a $2\times4$ configuration, with an active area of $8\times8$ mm$^2$ each, providing a total active area of 5.12 cm$^2$ per module with a 1 mm dead region along the device's borders. The detectors operate via electron drift in a fully depleted $n$-type silicon bulk, with charge collected at a small central anode. The $n^-$ silicon substrate is 450 \textmu m thick, providing a detection efficiency above 85$\%$ in the 4–12 keV energy range \cite{Miliucci:2021wbj}. A defining characteristic of SDDs is the small anode capacitance ($\approx$100 fF), resulting in reduced electronic noise and high energy and time resolutions, independent of the active detector area. Each SDD array is coupled to eight CMOS (Complementary Metal-Oxide Semiconductor) low-noise, pulsed reset, charge sensitive cryogenic pre-amplifier, named CUBE \cite{bombelli2010low,bombelli2011cube}, integrated on the ceramic carrier. Thanks to this pre-amplifier system, the SDDs' performance is stable even when exposed to high and variable particle rates. The output of the CUBE pre-amplifier is connected to an Application Specific Integrated Circuit (ASIC) called SFERA (SDDs Front-End Readout ASIC) \cite{Quaglia:2016uox,schembari2016sfera}, which performs analog shaping and peak detection of the signals.

\section{SDD calibration procedure and energy response}
This study characterizes, for the first time, the energy response of the SIDDHARTA-2 SDD system up to 50 keV, as required for future high-energy kaonic atom measurements \cite{Curceanu:2023yuy}. The calibration was performed by measuring the energies of the fluorescence lines emitted by several elements. To cover the energy range up to 50 keV, two data sets were used. The first set was acquired during a data taking with DA$\Phi$NE's beams in collision mode. This set of data exhibits fluorescence lines, arising from excitations of the materials of the setup, in the 10-30 keV energy range. The second set of data was specifically collected with the goal to test the goodness of the calibration procedure at $\sim50$ keV: this specific data taking was performed with a $\beta^-$ $^{90}$Sr source to excite the X-ray fluorescence lines of a $^{169}$Tm target. In Table \ref{tab:energies} the nominal values of the energies of all the transition lines used to perform this study are reported \cite{kortright2001x}.

\begin{table}[H]
    \centering
    \caption{Fluorescence X-ray lines used for calibration and their nominal energies \cite{kortright2001x}, following the IUPAC notation \cite{jenkins1991iupac}. With the exception of the Ba and Tm lines, the reported energies correspond to the intensity-weighted mean of the individual components contributing to each transition.}
    \label{tab:energies}
    \begin{tabular}{lcc}
    \hline
    Element & Transition & Energy [keV] \\
    \hline
    Bi  & L$_3$–M$_{5,4}$ (L$_{\alpha}$) & 10.828 \\
    Bi  & L$_2$–M$_{5,4}$ (L$_{\beta}$) & 13.009 \\
    Pd  & K–L$_{3,2}$ (K$_{\alpha}$) & 21.122 \\
    Ag  & K–L$_{3,2}$ (K$_{\alpha}$) & 22.103 \\
    Ba  & K–L$_3$ (K$_{\alpha_1}$) & 32.194 \\
    Tm  & K–L$_2$ (K$_{\alpha_2}$) & 49.773 \\
    Tm  & K–L$_3$ (K$_{\alpha_1}$) & 50.742 \\
    \hline
    \end{tabular}
\end{table}

\begin{figure}[H]
        \centering
        \includegraphics[width=0.80\textwidth]{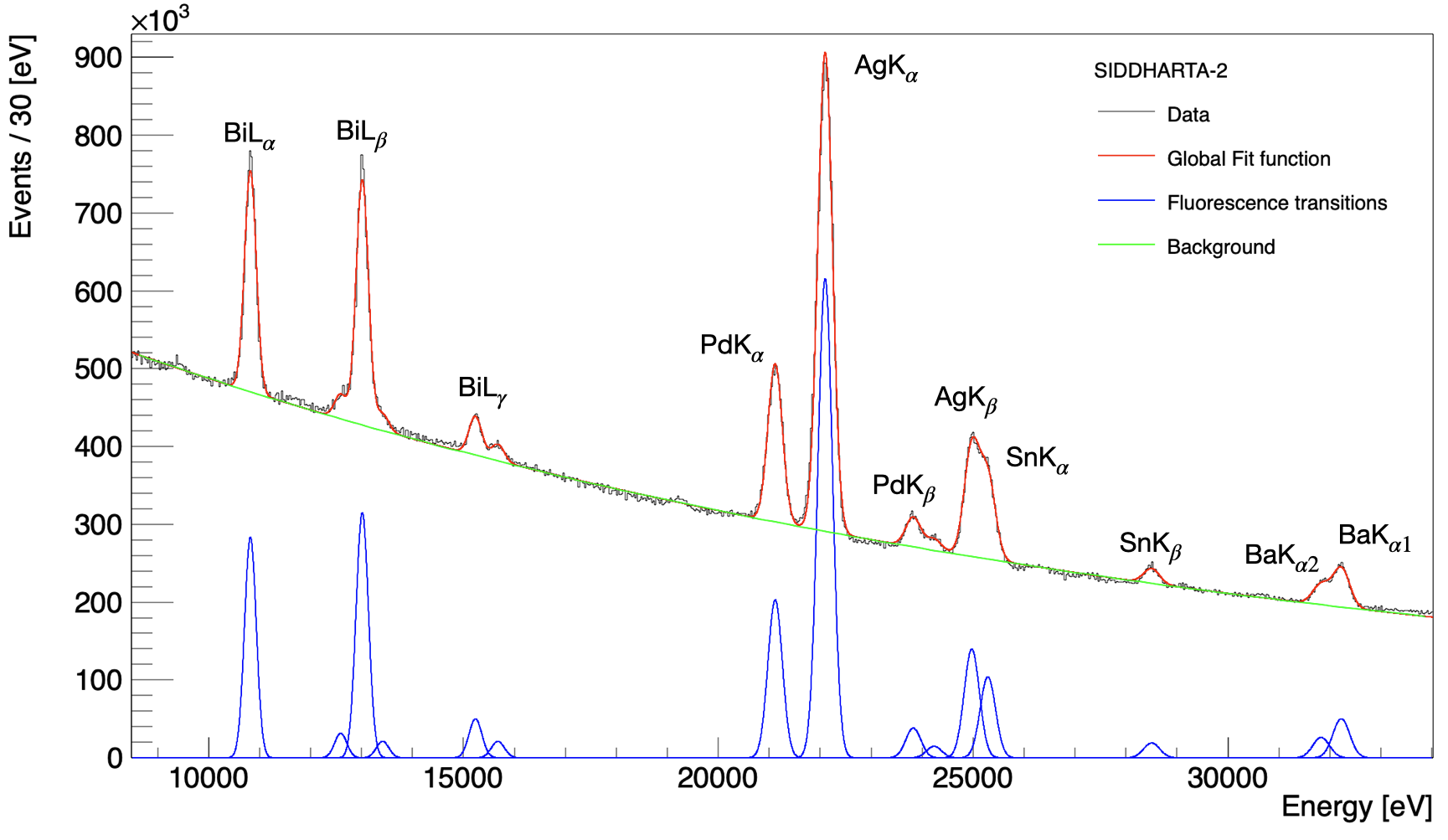}
        \includegraphics[width=0.80\textwidth]{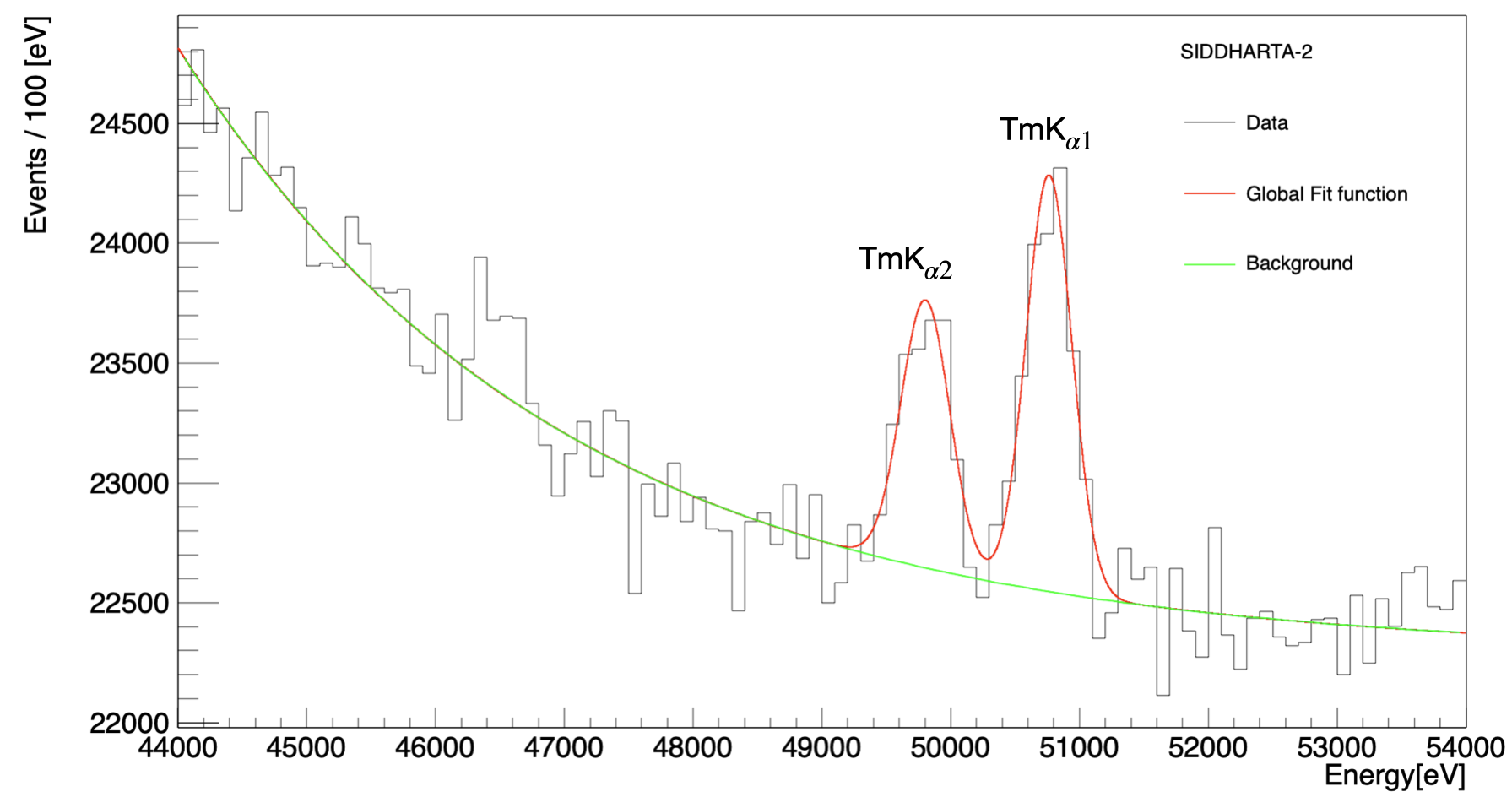}
        \caption{Above: fit of the energy spectrum for the run in beams collision mode, obtained by summing up all the SDDs' calibrated spectra. Below: fit of the calibrated energy spectrum obtained with a $^{169}$Tm target.}
        \label{fig:spectra}
\end{figure}

\noindent
The calibration of each SDD over this extended energy range is performed with a linear calibration function exploiting the transition lines with the highest statistics, namely BiL$_{\alpha}$, PdK$_{\alpha}$, and AgK$_{\alpha}$. The BaK$_{\alpha1}$, TmK$_{\alpha_2}$ and TmK$_{\alpha1}$ transition lines are then employed to assess the accuracy of the resulting energy calibration up to 50 keV. The first step of the procedure was a fit of the measured spectrum of each SDD for the run with beams in collision to extract the position, in ADC (Analog to Digital Converter) counts, of the transitions centroids and their FWHM. Each peak is described by a Gaussian function:
\begin{equation} \label{eq:gauss}
    G(x) = \frac{Gain}{\sqrt{2\pi}\sigma} e^{\frac{-(x-x_0)^2}{2\sigma^2} } \, ,
\end{equation}
where $x_0$ is the mean value of the Gaussian, $\sigma$ its standard deviation (width), and $Gain$ is a normalization factor proportional to the peak amplitude. This is summed to a tail function to account for the low-energy contributions \cite{Gysel:2003}:
\begin{equation} \label{eq:tail}
    T(x) = \frac{Gain}{2\beta\sigma} e^{\frac{(x-x_0)}{\beta\sigma} + \frac{1}{2\beta^2} } \mathrm{erfc} \left( \frac{x-x_0}{\sqrt{2}\sigma} + \frac{1}{\sqrt{2}\beta} \right) \, ,
\end{equation}
where $\beta$ is the slope parameter of the tail function. All parameters were left free in the fit. The FWHM of each peak increases with energy according to the relation: 
\begin{equation} \label{eq:fwhm}
    \mathrm{FWHM}(E) = 2\sqrt{2\ln2} \cdot \sqrt{\varepsilon \mathrm{F}E + \left(\frac{\mathrm{N}}{2\ln2}\right)^2} \, ,
\end{equation}
where $\varepsilon$ is the mean energy required to create an electron–hole pair in silicon ($\sim$3.71 eV), while $\mathrm{F}$ and $\mathrm{N}$ represent the Fano factor and the intrinsic electronic noise of the SDDs, respectively. The energy calibration of each detector is obtained by fitting the measured peak positions with a first-order polynomial function:
\begin{equation}
    E_{\mathrm{meas}} = g \cdot \mathrm{ADC}_{\mathrm{meas}} + c \, ,
\end{equation} \label{eq:calib}
where $\mathrm{ADC}_{\mathrm{meas}}$ represents the measured centroid position of each calibration peak in ADC channels, $g$ is the calibration gain and $c$ the intercept. After calibration, the spectra of all SDDs are summed to obtain the overall energy-calibrated spectrum shown in Figure \ref{fig:spectra}.

\begin{figure}[H]
    \centering
    \includegraphics[width=0.80\textwidth]{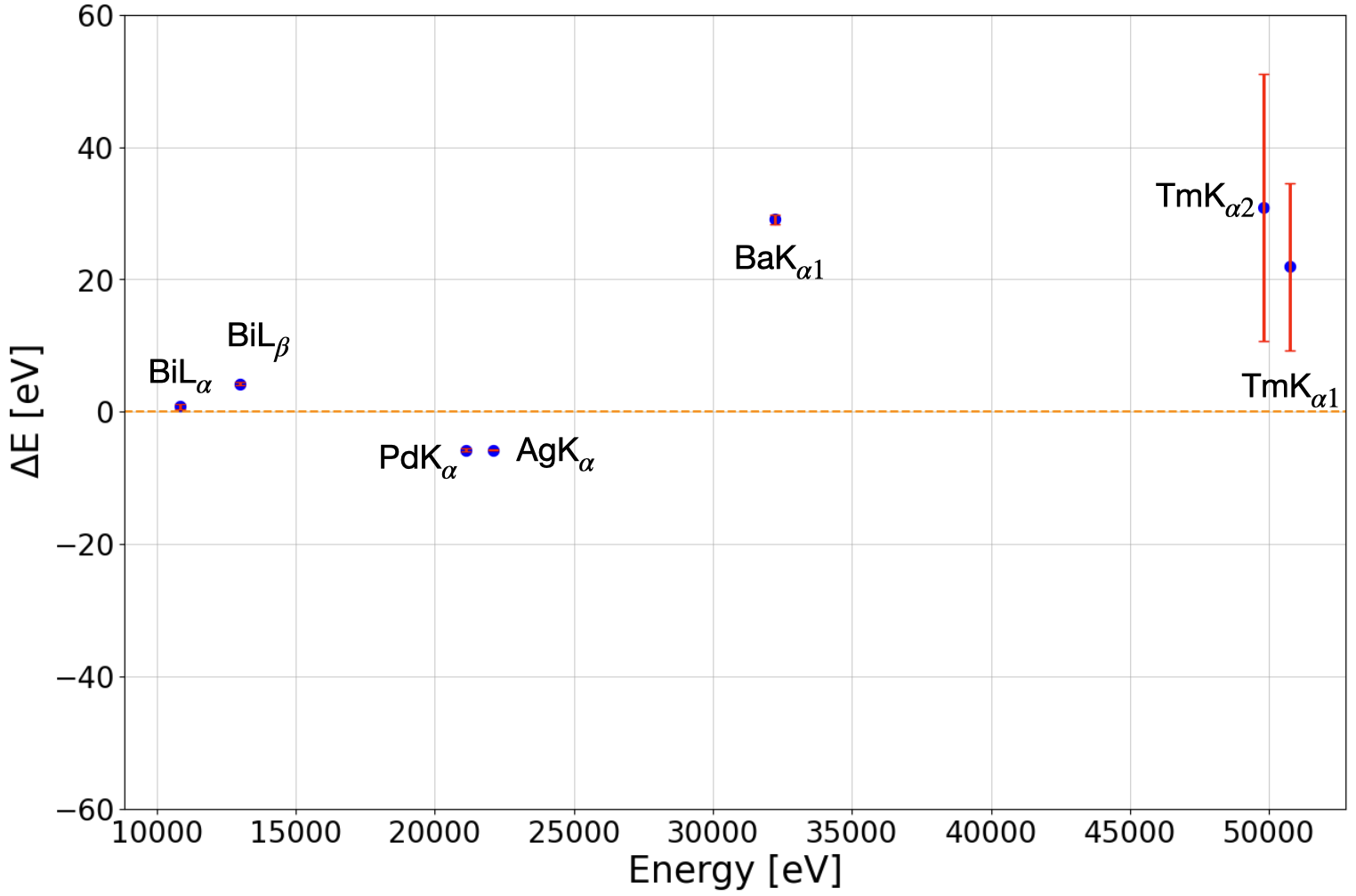}
    \caption{Residuals of the X-ray transitions of interest with the calibration performed with a linear fitting function.}
    \label{fig:residuals}
\end{figure}

\noindent
The residuals of the measured energies ($E_{meas}$) of the transition lines from their nominal values ($E_{th}$) (see Table \ref{tab:energies}) were extracted to evaluate the linearity of the SDDs' energy response and their relative accuracy. The plot of the energy residuals, defined as $\Delta E = (E_{meas}-E_{th})$, for the transition lines with the highest statistics and for the Ba and Tm transitions is reported in Figure \ref{fig:residuals} and their values are presented in Table \ref{tab:residuals}. The errors correspond to the fit uncertainties on the peak positions. The larger error bars on the K$_{\alpha1}$ and K$_{\alpha2}$ lines of Tm are due to the less statistics acquired during the data taking with the $^{169}$Tm target. The SDDs show a linear energy response in the 10–50 keV range with relative precision $\Delta E/E < 10^{-3}$.

\begin{table}[H]
    \centering
    \caption{Residuals of the measured energies from their nominal values (see Table \ref{tab:energies}).}
    \begin{tabular}{cc}
        \hline
        \textbf{Transition} & \textbf{Residual $\Delta$E [eV]} \\
        \hline
        BiL$_{\alpha}$ & 0.79 $\pm$ 0.43 \\
        BiL$_{\beta}$ & 4.22 $\pm$ 0.20 \\
        PdK$_{\alpha}$ & -5.83 $\pm$ 0.23 \\
        AgK$_{\alpha}$ & -5.81 $\pm$ 0.09 \\
        BaK$_{\alpha_1}$ & 29.10 $\pm$ 0.73 \\
        TmK$_{\alpha_2}$ & 30.80 $\pm$ 20.19 \\
        TmK$_{\alpha_1}$ & 21.90 $\pm$ 12.64 \\
        \hline
    \end{tabular}
    \label{tab:residuals}
\end{table}

\noindent
To demonstrate the goodness of the achieved calibration accuracy for future kaonic atoms measurements, Table \ref{tab:EXK_transitions} reports the expected energies, strong-interaction shifts and widths of selected kaonic atoms (Li, Be, B) on transitions in the 15-50 keV energy range targeted by EXKALIBUR \cite{Curceanu:2023yuy,Curceanu:2026zjg}. The expected strong-interaction shifts, of order 30-200 eV, are well above the achieved energy-scale precision in the 10-50 keV range, allowing precise measurements of these transitions in future studies of kaonic atoms.

\begin{table}[H]
\centering
\caption{Energies, strong-interaction widths and shifts of selected kaonic atoms $2p$ transitions targeted by the EXKALIBUR experiment \cite{Curceanu:2023yuy}.}
\label{tab:EXK_transitions}
\begin{tabular}{lccc}
\hline
Isotope & E$_{3d\to2p}$ [keV] & $\varepsilon_{2p}$ [eV] & $\Gamma_{2p}$ [eV] \\
\hline
$^{6}$Li  & 15.33  & $-2.8$   & 44   \\
$^{7}$Li  & 15.33  & $-3.2$   & 54   \\
$^{8}$Be  & 27.71  & $-34.8$  & 225  \\ 
$^{9}$Be  & 27.71  & $-36.8$  & 247  \\
$^{10}$B  & 43.568 & $-170$   & 718  \\
$^{11}$B  & 43.768 & $-179$   & 757  \\
\hline
\end{tabular}
\end{table}

\begin{figure}[H]
    \centering
    \includegraphics[width=0.80\textwidth]{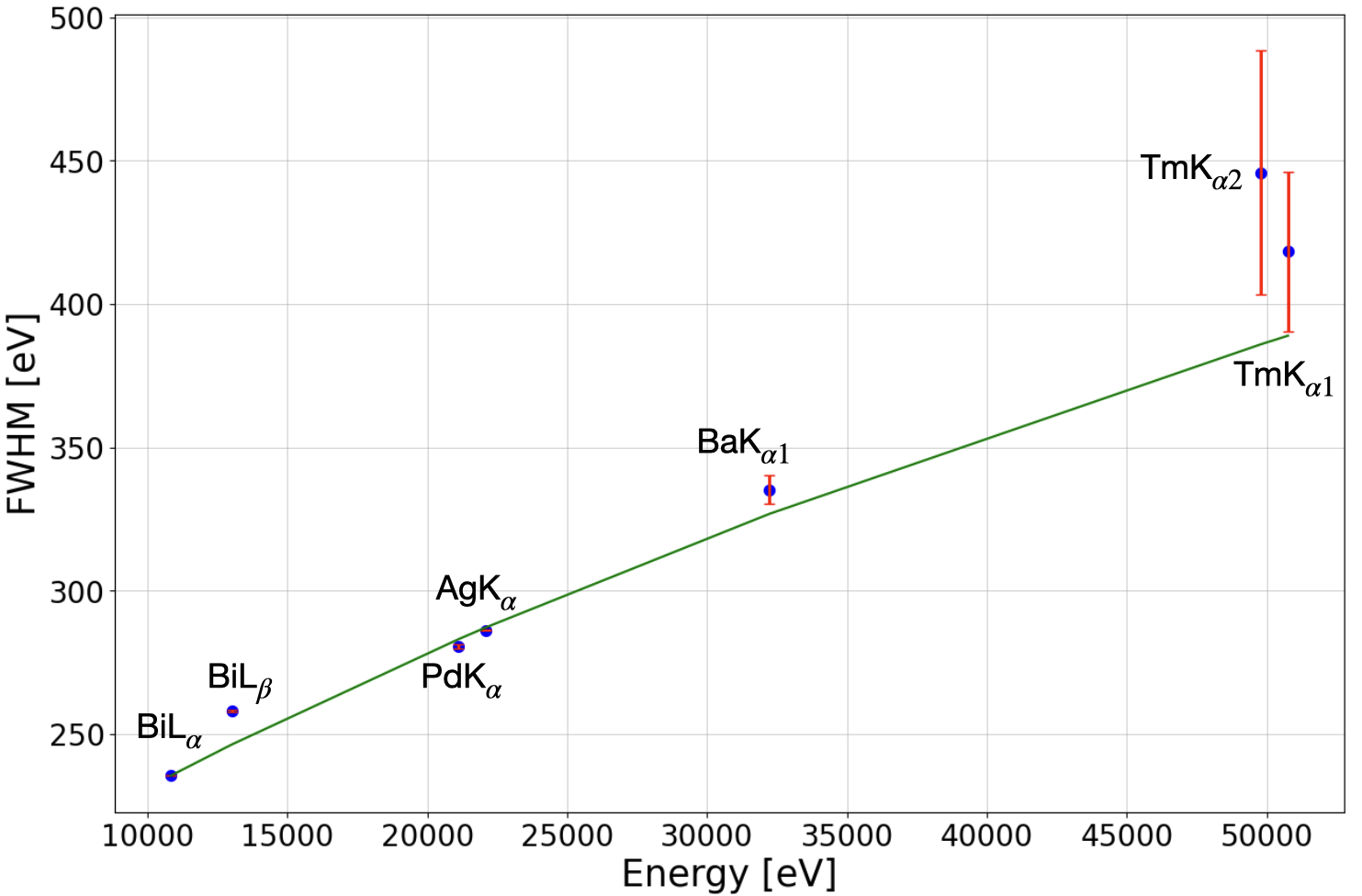}
    \caption{FWHM of the measured X-ray transition lines as function of the energy obtained with the linear calibration.}
    \label{fig:fwhm}
\end{figure}

\noindent
Finally, the energy resolution of the SDDs is evaluated as the FWHM of the detected X-ray peaks. This is obtained from the standard deviation $\sigma$ of the Gaussian component of the fitted peaks as FWHM $ = 2\sqrt{2\ln2} \, \sigma$. The uncertainties on the FWHM values are propagated directly from the fit errors of $\sigma$. The dependence of the FWHM on energy is shown in Figure \ref{fig:fwhm}, together with a fit to the data based on Equation \ref{eq:fwhm}, where $F$ and $N$ were treated as free parameters. The best-fit values are $\mathrm{F} = 0.116579 \pm 0.000068$ and $\mathrm{N} = (172 \pm 29)$~eV. When compared with the strong–interaction widths of the kaonic atoms $2p$ transitions listed in Table \ref{tab:EXK_transitions}, which range from a few tens of eV for kaonic lithium up to several hundred eV for kaonic boron, the achieved energy resolution in the 10–50 keV range is fully adequate for resolving and extracting these widths.

\begin{table}[H]
    \centering
    \caption{FWHM of the measured X-ray transition lines.}
    \label{tab:FWHM}
    \begin{tabular}{cc}
        \hline
        \textbf{Transition} & \textbf{FWHM [eV]} \\
        \hline
        BiL$_{\alpha}$ & 235.50 $\pm$ 0.01\\
        BiL$_{\beta}$ & 257.97 $\pm$ 0.40\\
        PdK$_{\alpha}$ & 280.60 $\pm$ 0.75\\
        AgK$_{\alpha}$ & 286.26 $\pm$ 0.27\\
        BaK$_{\alpha_1}$ & 335.3 $\pm$ 5.1\\
        TmK$_{\alpha_2}$ & 446 $\pm$ 43\\
        TmK$_{\alpha_1}$ & 418 $\pm$ 28\\
        \hline
    \end{tabular}
\end{table}

\section{Conclusions}
The SIDDHARTA-2 SDD system has been characterized over an extended energy range up to 50 keV, confirming its suitability for future high-precision X-ray spectroscopy of kaonic atoms \cite{Curceanu:2023yuy}. The linear calibration function provides a relative energy precision better than $10^{-3}$ across the 10–50 keV range, validating the linearity and stability of the system. The achieved energy resolution and calibration accuracy allow precise and reliable measurements of the expected strong-interaction shifts and widths of kaonic atom transitions in the EXKALIBUR program (Table \ref{tab:EXK_transitions}).\\
These results demonstrate that the SIDDHARTA-2 SDD system is capable of performing precision measurements of kaonic lithium, beryllium, and boron X-ray transitions, providing a unique tool to study kaon–multi-nucleon interactions and bound-state QED effects in few-body systems. The achieved performance opens the way for extending the kaonic atom program beyond light systems, enabling systematic studies of intermediate-mass and heavy kaonic atoms with transition energies up to 50 keV.
The demonstrated linearity, energy resolution, and calibration accuracy ensure reliable measurements of strong-interaction-induced energy shifts and broadenings in future kaonic atom studies.

\section*{Acknowledgments}
We thank C. Capoccia from LNF-INFN and H. Schneider, L. Stohwasser, and D. Pristauz-Telsnigg from Stefan Meyer-Institut for their fundamental contribution in designing and building the SIDDHARTA-2 setup. We thank as well the DA$\Phi$NE staff for the excellent working conditions and permanent support. Special thank to Catia Milardi for her continued support and contribution during the data taking. Part of this work was supported by the Austrian Science Fund (FWF): [P24756-N20 and P33037-N] and FWF Doctoral program No. W1252-N27; the Croatian Science Foundation under the project IP-2022-10-3878; the EU STRONG-2020 project (Grant Agreement No. 824093); the EU Horizon 2020 project under the MSCA (Grant Agreement 754496); the Japan Society for the Promotion of Science JSPS KAKENHI Grant No. JP18H05402; the Polish Ministry of Science and Higher Education grant No. 7150/E-338/M/2018 and the Polish National Agency for Academic Exchange( grant no PPN/BIT/2021/1/00037); the EU Horizon 2020 research and innovation programme under project OPSVIO (Grant Agreement No. 101038099); Project 963530 - RCLIB, "Recycling of the cathodes, based on carbon nanotubes and conducting polymers, from spent rechargeable Li batteries" - call LEAP-RE 2022. The authors acknowledge support from the SciMat and qLife Priority Research Areas budget under the program Excellence Initiative—Research University at the Jagiellonian University. Catalina Curceanu acknowledge University of Adelaide, where part of this work was done (under the George Southgate fellowship, 2024).

\section*{References}
\bibliography{iopart-num}

\end{document}